\documentclass[twocolumn,showpacs,preprintnumbers,prl]{revtex4}
\usepackage{amsmath,amsfonts,amssymb,braket,graphicx}
\usepackage{hyperref}
\usepackage{setspace}
\usepackage[english]{babel}
\usepackage{graphicx}
\usepackage{dcolumn}
\usepackage{bm}
\usepackage{dsfont}
\usepackage{epstopdf}
\usepackage{gensymb}
\usepackage{nicefrac} 
\usepackage{color}

\newcommand{\M}{\mathcal M}
\newcommand{\E}{\mathcal E}

\def\1{\mathchoice{\rm 1\mskip-4.2mu l}{\rm 1\mskip-4.2mu l}{\rm
		1\mskip-4.6mu l}{\rm 1\mskip-5.2mu l}}
 
\begin{document}
 
\twocolumngrid
 
\title{Experimental test  of  entropic noise-disturbance uncertainty relations \\ for  spin-$\nicefrac{1}{2}$ measurements}

\author{Georg Sulyok$^{1}$}
\author{Stephan Sponar$^1$}
\author{B\"{u}lent Demirel$^1$}
\author{Francesco Buscemi$^2$}
\author{Michael J. W. Hall$^3$}
\author{Masanao Ozawa$^2$}
\author{Yuji Hasegawa$^{1}$}

\affiliation{$^1$Atominstitut, Vienna University of Technology, 1020 Vienna, Austria}
\affiliation{$^2$Graduate School of Information Science, Nagoya University, Chikusa-ku, Nagoya 464-8601, Japan}
\affiliation{$^3$Centre for Quantum Computation and Communication Technology	(Australian Research Council), Centre for Quantum Dynamics, Griffith University, Brisbane, QLD 4111, Australia}

\date{\today}

\begin{abstract}
Information-theoretic definitions for noise and disturbance in quantum measurements were given in [Phys.\ Rev.\ Lett.\ \textbf{112}, 050401 (2014)] and a state-independent noise-disturbance uncertainty relation was obtained. Here, we derive a tight noise-disturbance uncertainty relation for complementary qubit observables and carry out an experimental test. Successive projective measurements on the neutron's spin-$\nicefrac{1}{2}$ system, together with a correction procedure which reduces the disturbance, are performed. Our experimental results saturate the tight noise-disturbance uncertainty relation for qubits when an optimal correction procedure is applied.
\end{abstract}
  

\maketitle

\emph{Introduction -}
The uncertainty principle, first formulated by Heisenberg in 1927 \cite{Heisenberg27}, expresses an intuitive understanding of the physical consequences of non-commutativity.
Heisenberg argued that it is impossible to simultaneously measure noncommuting observables with arbitrary precision, and used the famous $\gamma$-ray microscope thought-experiment to obtain the 
`noise-disturbance' uncertainty relation
\begin{equation} \label{heisenberg}
q_1\, p_1 \sim h
\end{equation}
for the product of the mean error (noise) $q_1$ of a position measurement and the discontinuous change (disturbance) $p_1$ of the particle's momentum. 
In the subsequent mathematical derivation of Eq.~(\ref{heisenberg}),
he showed that the product of the position and momentum standard deviations, $\Delta q\,\Delta p$, 
was equal to $\hbar/2$ for a class of Gaussian wavefunctions, which was generalised by Kennard to 
\begin{equation} \label{kennard}
\Delta q\,\Delta p\geq \frac{\hbar}{2}
\end{equation} 
for all states \cite{Kennard27}. 

Note that relation (\ref{kennard}) sets a limitation as to how one can precisely prepare 
both the position and momentum of a quantum system, 
independently of whether these observables are actually measured.  
Hence, such preparation relations, whether formulated in terms of standard deviations \cite{Kennard27,Robertson29} or Shannon entropies \cite{Deutsch_uncertainty}, do not place any restrictions {\it per se} on joint or successive measurements of noncommuting observables.

In contrast, Heisenberg's formulation of the uncertainty relation in Eq.~(\ref{heisenberg}) is all about the unavoidable influence of measuring instruments on quantum systems: the more precisely one observable such as position is measured,  
the greater is the  disturbance to another observable such as momentum \cite{Heisenberg27,Heisenberg30}.
 As an aside, we notice that Eq.~(\ref{heisenberg}) can in fact be derived from Eq.~(\ref{kennard}) under the repeatability hypothesis \cite{Ozawa14}, which was implicitly assumed in most of arguments on quantum measurements until the 1970's \cite{Davies70}. 
 
 A rigorous error-disturbance uncertainty relation, generalising Eq.~(\ref{heisenberg}) to arbitrary pairs of observables 
and measurements without assuming the repeatability hypothesis, 
was derived by Ozawa \cite{Ozawa03, OzawaPLA03, Ozawa04} 
and has recently received considerable attention. 
The validity of Ozawa's relation, as well as of a stronger version of this relation \cite{Branciard13}, 
were experimentally tested with neutrons \cite{Erhart12, Sulyok13,Sponar14} 
and with photons \cite{Steinberg12, Edamatsu13,Hall13, Kaneda14, Ringbauer14}. 
Other approaches generalising Heisenberg's original relation 
can be found, for example,  in \cite{Busch13,Busch14,Lu14}.

It is very natural to also seek a formulation of the uncertainty principle in terms of the information gained and lost due to measurement influences. Such a formulation was recently introduced by  Buscemi \emph{et al.} \cite{Buscemi14}, leading to a state-independent uncertainty relation.  Here, noise and disturbance are quantified not  by a difference between a system observable and the quantity actually measured, but by the correlations between input states and measurement outcomes, independently of how they are labelled. In this letter we derive a tight uncertainty relation for information-theoretic noise and disturbance, in the qubit case, and demonstrate its validity in a neutron polarimeter experiment.

\begin{figure}[!t]
  	\includegraphics[width=85mm]{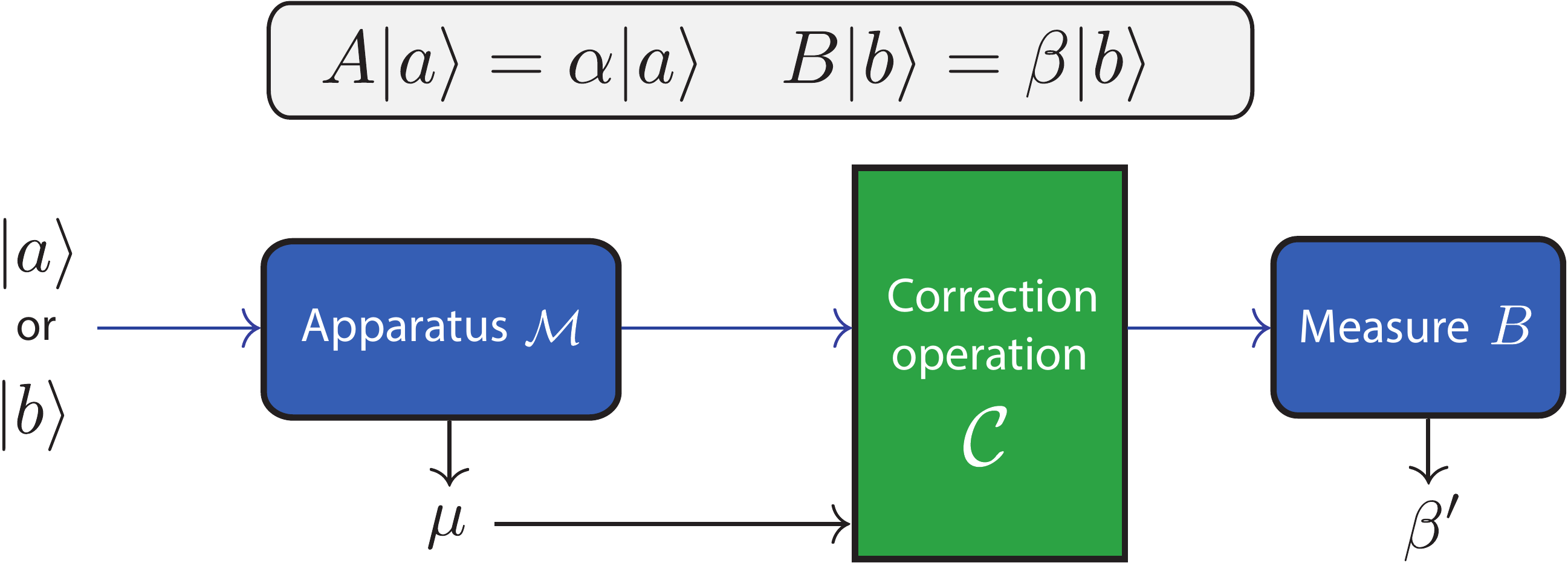}
  	\caption{(color online) Experimental concept for determination of noise and disturbance. Randomly selected eigenstates of $A$ and $B$ are sent into a measurement apparatus $\mathcal M$. After a correction operation $\mathcal C$, and a precise measurement of $B$, the information-theoretic noise $N(\mathcal M, A)$ and disturbance $D(\mathcal M, B)$ are calculated using the conditional probabilities $p(\alpha|\mu)$ and $p(\beta|\beta^\prime)$, respectively.\label{fig:scheme}}
\end{figure}

\emph{Theoretical framework -} Consider an observable $A$, acting on a finite-dimensional Hilbert space, with eigenvalues $\alpha$ belonging to the non-degenerate eigenstates $\ket{a}$  and a measurement apparatus $\mathcal M$ representing a quantum instrument \cite{davies1970,davies_book,Ozawa_JMathPhys_1984} with possible outcomes $\mu$. All eigenstates $\ket{a}$ of $A$ are now fed with equal probability into the apparatus, which is schematically illustrated in Fig.\,{\ref{fig:scheme}}. The conditional probability $p(\alpha|\mu)$ that the eigenstate $\ket{a}$ was sent, given a specific measurement outcome $\mu$, and the marginal probability $p(\mu)$ for occurrence of the specific outcome are used to define the information-theoretic noise $N(\mathcal M, A)$ as
\begin{equation}
\label{eq:noise_definition}
N(\mathcal M, A) := - \sum_{\alpha,\mu} p(\mu) p(\alpha|\mu) \log p(\alpha|\mu)= H(\mathbb{A}|\mathbb{M}).
\end{equation}
Equation\,(\ref{eq:noise_definition}) is just the conditional entropy $H(\mathbb{A}|\mathbb{M})$, where $\mathbb{A}$ and $\mathbb{M}$ denote the classical random variables associated with input $\alpha$ and output $\mu$. The information-theoretic noise thus quantifies how well the value of $A$ can be inferred from the measurement outcome and only vanishes if an absolutely correct guess is possible.

The information-theoretic disturbance is defined in a similar manner as
\begin{equation}
\label{eq:disturbance_definition}
D(\mathcal M, B) := - \sum_{\beta,\beta'} p(\beta') p(\beta|\beta') \log p(\beta|\beta')= H(\mathbb{B}|\mathbb{B}').
\end{equation}
Here uniformly distributed eigenstates $\ket{b}$ of an observable $B$ are input to the apparatus  $\mathcal M$, and a subsequent measurement of $B$ is performed, with outcomes labeled by $\beta'$ (Fig.\,{\ref{fig:scheme}}). The disturbance $D(\mathcal M, B)$ thus quantifies the correlation between the initial and final values of $B$, and is a measure of how much information about $B$ is lost through the measurement $\mathcal M$. 

In order to determine the \emph{irreversible} loss of information about $B$, a correction operation $\mathcal{C}$ can be performed before the $B$-measurement to decrease the disturbance (Fig.\,{\ref{fig:scheme}}), and consists of any completely positive, trace preserving map. We deal with two cases here; one is the uncorrected disturbance which we write as $D_0$. The other is the optimally corrected disturbance denoted as $D_{\rm opt}$ corresponding to the correction operation that minimizes the disturbance. For any correction procedure the information theoretic noise and disturbance fulfil the following uncertainty relation \cite{Buscemi14}
\begin{equation} 
\label{eq:ndur}
N(\mathcal M,A) + D(\mathcal M,B) \geq 
c_{AB}:=- \log\max  |\langle  a |  b\rangle |^2, 
\end{equation}
where $\ket{  a}$ and $\ket{  b}$ denote the eigenstates of the observables $A$ and $B$. 

For maximally incompatible qubit observables, represented by the Pauli matrices $\sigma_z$ and $\sigma_y$, we have been able to significantly strengthen this relation (see Sec.\,I of the Supplemental Material \cite{supp_entropic_PRL}) to the tight relation
\begin{equation}
\label{eq:qubits_ndur}
g[N(\mathcal M,\sigma_z)]^2+g[D(\mathcal M,\sigma_y)]^2\leq 1.
\end{equation}
Here $g[x]$ denotes the inverse of the function $h(x)$ on the interval $x \in [0,1]$ given by
\begin{equation}
\label{eq:h_def}
h(x):=-\frac{1+x}{2}\log \frac{1+x}{2} - \frac{1-x}{2}\log \frac{1-x}{2}.
\end{equation}

\begin{figure}[!b]
  	\includegraphics[width=84mm]{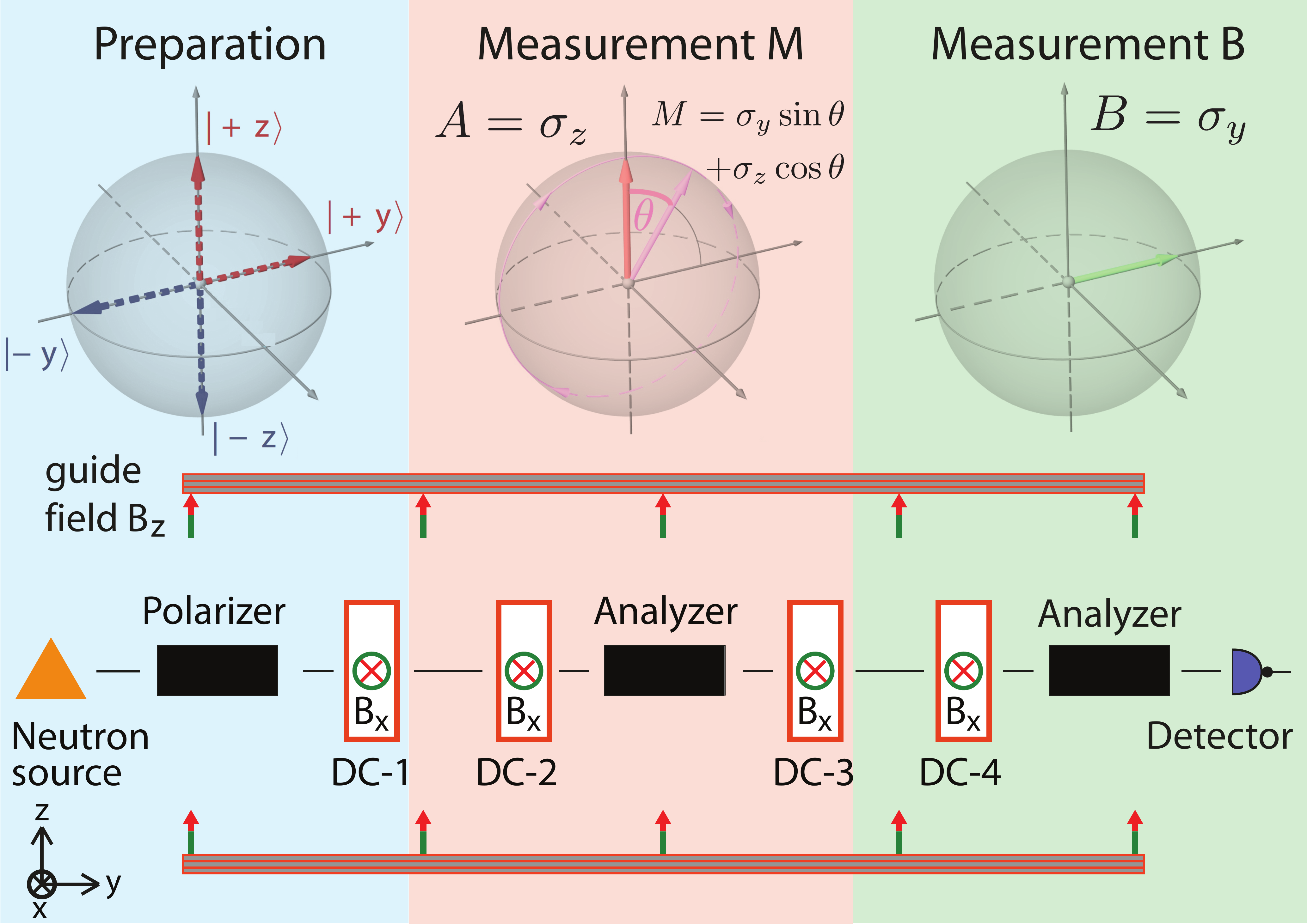}
  	\caption{(color online) Neutron polarimetric setup for the demonstration of information-theoretic uncertainty relations for noise and disturbance. Exploiting Larmor precession of the Bloch vector around magnetic fields ($B_x$, $B_z$) and using supermirror arrays (polarizer, analyzers) as projectors all required spin states can be prepared  and measured. }\label{fig:setupUnCorr}
\end{figure}

{\emph{Experimental procedure -} In our experiment projective measurements on neutron spin qubits are utilized. The observables are chosen to be Pauli spin matrices $A=\sigma_z$ and $B=\sigma_y$, having the eigenvalues $\alpha=\pm 1$ and $\beta=\pm 1$. We denote the eigenstates of $A$ and $B$ as $\ket{\alpha z} \equiv \ket{\pm
z}$ for $\alpha = \pm 1$ and $\ket{\beta y} \equiv \ket{\pm y}$ for $\beta=\pm 1$, respectively. For projective measurements the measurement apparatus $\mathcal M$ is simply characterized by a measurement operator
\begin{equation}
\label{eq:M}
M=\vec m\cdot\vec{\sigma} = \sigma_y \sin\theta + \sigma_z\cos\theta,
\end{equation}
representing spin along the axis $\vec m =(0,\sin\theta,\cos\theta)$. It has the eigenvalues/outcomes $\mu=\pm 1$ and projects the system onto its eigenstates  denoted as $\ket{\mu m} \equiv\ket{\pm m}$ for $\mu = \pm 1$ after the measurement.

The experiment is performed on the neutron's spin-$\nicefrac{1}{2}$ qubit system using the polarimeter beam line of the tangential beam port at the research reactor facility TRIGA Mark II of the Vienna University of Technology \cite{Erhart12,Sulyok13}. The setup is depicted in Fig.\,\ref{fig:setupUnCorr} and illustrates the generic experimental procedure: An unpolarized thermal neutron beam, incident from a pyrolytic graphite crystal, with a mean wavelength of 2.02 \AA\, and spectral width $\Delta\lambda/\lambda=0.015$, is spin polarized up to $\sim99\,\%$ via reflection from a bent Co-Ti supermirror array, with polarization in the $+z$-direction. To prevent depolarization by stray fields a 13\,Gauss guide field $B_z$ pointing in the $+z$-direction is applied along the entire setup. 

For the generation of the desired initial states $\ket{\pm z}$ and $\ket{\pm y}$ the first spin turner coil DC-1 is used. Within the coil region a field $B_x$, pointing in $+x$-direction, is effectively applied. Larmor precession around the x-axis is induced and the strength of $B_x$ is tuned such that it causes a spin rotation by an angle of 0, $\pi$, or $\pm\pi/2$ radians within the coil DC-1. In order to achieve the uniform distribution of the eigenstates as required for the determination of noise and disturbance all four input states are sent one after another.

For the measurement of $M$ another spin turner coil (DC-2) is used. It is placed such that within the distance to DC-1 integer multiples of the full rotation period around the z-axis are performed in the guide field. Then, by correctly adjusting the strength of $B_x$ in DC-2, the $\ket{\mu m}$-component of the spinor is rotated to $\ket{+z}$. After the projection onto $\ket{+z}$ in the second supermirror (first analyzer in Fig.\,\ref{fig:setupUnCorr}) spin turner coil DC-3 rotates the analyzer's output state $\ket{+z}$ to $\ket{\mu m}$ thus completing the projective measurement of $M$.
In an analogous way, DC-4 and the third supermirror perform the $B$-measurement. The recovering of the eigenstates of $B$ can be omitted since the neutron detector is not sensitive to spin (for more details of the experimental procedure see Sec.\,II of \cite{supp_entropic_PRL}).

The two successively performed projective spin measurements result in four output intensities for each input eigenstate. We label the intensities as $I^{A}_{\alpha \mu \beta'}$ and $I^{B}_{\beta \mu \beta'}$ where all lower indices can take the values $\pm 1$. The different upper indices $A$ and $B$ discriminate between the input states. For example, $I^{A}_{+\, .\, .}$ indicates that the eigenstate $\ket{+ z}$ of $A =\sigma_z$ has been sent and $I^{B}_{+\, .\, .}$ stands for output intensities when $\ket{+ y}$ has been fed to the measurement apparatus. From $I^{A}_{\alpha \mu \beta'}$, the probabilities required for the determination of the information theoretic noise can be deduced, and  $I^{B}_{\beta \mu \beta'}$ yields the probabilities for the information theoretic disturbance:
\begin{eqnarray}
\label{eq:prob_noise_exp}
p(\alpha)&=&\frac{\sum_{\mu,\beta'} I^{A}_{\alpha \mu \beta'}}{\sum_{\alpha,\mu,\beta'} I^{A}_{\alpha \mu \beta'}}  \qquad 
p(\mu|\alpha) = \frac{\sum_{\beta'} I^{A}_{\alpha \mu \beta'}}{\sum_{\mu,\beta'} I^{A}_{\alpha \mu \beta'}}  \\
\label{eq:prob_dist_exp}
p(\beta)&=&\frac{\sum_{\mu,\beta'} I^{B}_{\beta \mu \beta'}}{\sum_{\beta,\mu,\beta'} I^{B}_{\beta \mu \beta'}}  \qquad 
p(\beta'|\beta) = \frac{\sum_{\mu} I^{B}_{\beta \mu \beta'}}{\sum_{\mu,\beta'} I^{B}_{\beta \mu \beta'}}
\end{eqnarray}
It is important to note here that we first determine which eigenstate has been sent and then record the probability for a specific outcome $\mu$. We thus obtain the conditioned probabilities $p(\mu|\alpha)$ and $p(\beta'|\beta)$ respectively instead of $p(\alpha|\mu)$ and $p(\beta|\beta')$ for which noise and disturbance are defined. But, by using Bayes' theorem for conditioned probabilities they can be converted into each other (see Sec.\,III of \cite{supp_entropic_PRL}).

\begin{figure}[!t]
		\includegraphics[width=70mm]{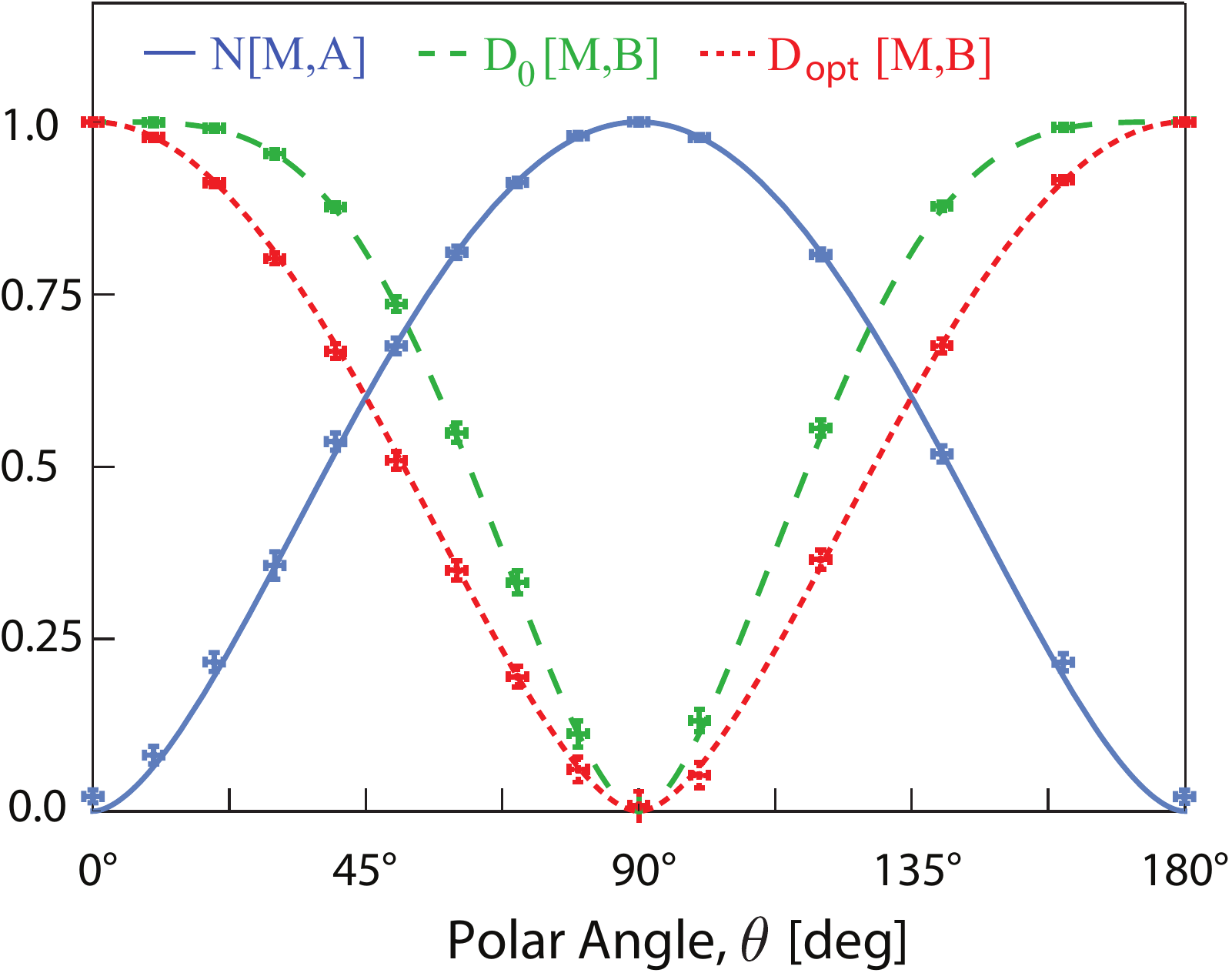}	
	\caption{(color online) Noise $N(\mathcal M,A)$ (straight blue line), uncorrected disturbance $D_{0}(\mathcal M,B)$ (dashed green line), and  optimally corrected disturbance $D_{\rm opt}(\mathcal M,B)$ (dotted red line)  vs. polar angle $\theta$ of $M$.
	\label{fig:NoiseDistUnCorr}}
\end{figure}

\emph{No correction procedure-}  In the first experiment, no additional correction is applied ($\mathcal{C}= \1$), we just successively measure $M$ and $B$. The setup then consists of the three stages depicted in Fig.\,\ref{fig:setupUnCorr}: i) state preparation - the corresponding eigenstates of the observables $\sigma_z$ and $\sigma_y$ are generated. ii) measurement of  $M=\sigma_y \sin\theta + \sigma_z\cos\theta$, and iii) measurement of $B=\sigma_y$.

We record the intensities while varying the polar angle $\theta$ of $M$ in the interval $\theta \in [0\degree,90\degree]$ with increment $\Delta \theta = 10\degree$ and with a smaller step width of $\Delta \theta = 20\degree$ in interval $\theta \in [100\degree,180\degree]$ since noise and disturbance are mirror-symmetric around $\theta =90\degree$. Fig.\,\ref{fig:NoiseDistUnCorr} shows the measured data points and their theory curves, with the latter given in terms of $h$ from Eq.\,(\ref{eq:h_def}) by 
\begin{equation}
\label{eq:noise_dist_qbit}
N(\mathcal M,A) = h\left(\cos\theta\right), \quad 
D_{0}(\mathcal M,B) =  h\left(\sin^2\theta\right).
\end{equation} 
An intuitive understanding for the information-theoretic meaning of noise and disturbance can be reached by looking at special values of $\theta$. For $\theta=0\degree$, the measurement operator $M=\sigma_z= A$. The measurement result $\mu$ is numerically identical to the "value" $\alpha$ of the observable $A$ in the eigenstate $\ket{\alpha z}$ and thus obviously perfectly correlated to it leading to vanishing noise. With increasing polar angle $\theta$ of $M$ the correlation is lost. For $\theta=90\degree$ ($M=\sigma_y$) the measurement outcome does not allow any inference as to which eigenstate of $A$ was sent, $p(\alpha|\mu)=\frac{1}{2}, \, \forall \alpha,\mu$. Thus, the information-theoretic noise is maximal. For $\theta=180\degree$ ($M=-\sigma_z$) the measurement result is perfectly anti-correlated with the observable's value. A guessing function of type $\alpha=f(\mu)=-\mu$ allows a flawless determination of the observable's value from the measurement outcome and thus, this setup is noiseless as well, although the numerical deviation between the measurement outcome and the value of $A$ is maximal. While in the standard "noise operator" approach \cite{OzawaPLA03,Ozawa04} the noise then becomes maximal (see for example error (dashed blue line) in Fig.\,8 of \cite{Sulyok13}), in the information-theoretic approach the degree of correlation rather than its sign is relevant. For the same reason, the behaviour of noise for $\theta= 180\degree$ to $360\degree$ is identical to $\theta= 0\degree$ to $180\degree$ (see also Sec.\,III of \cite{supp_entropic_PRL}). 

\begin{figure}[!b]
	\includegraphics[width=85mm]{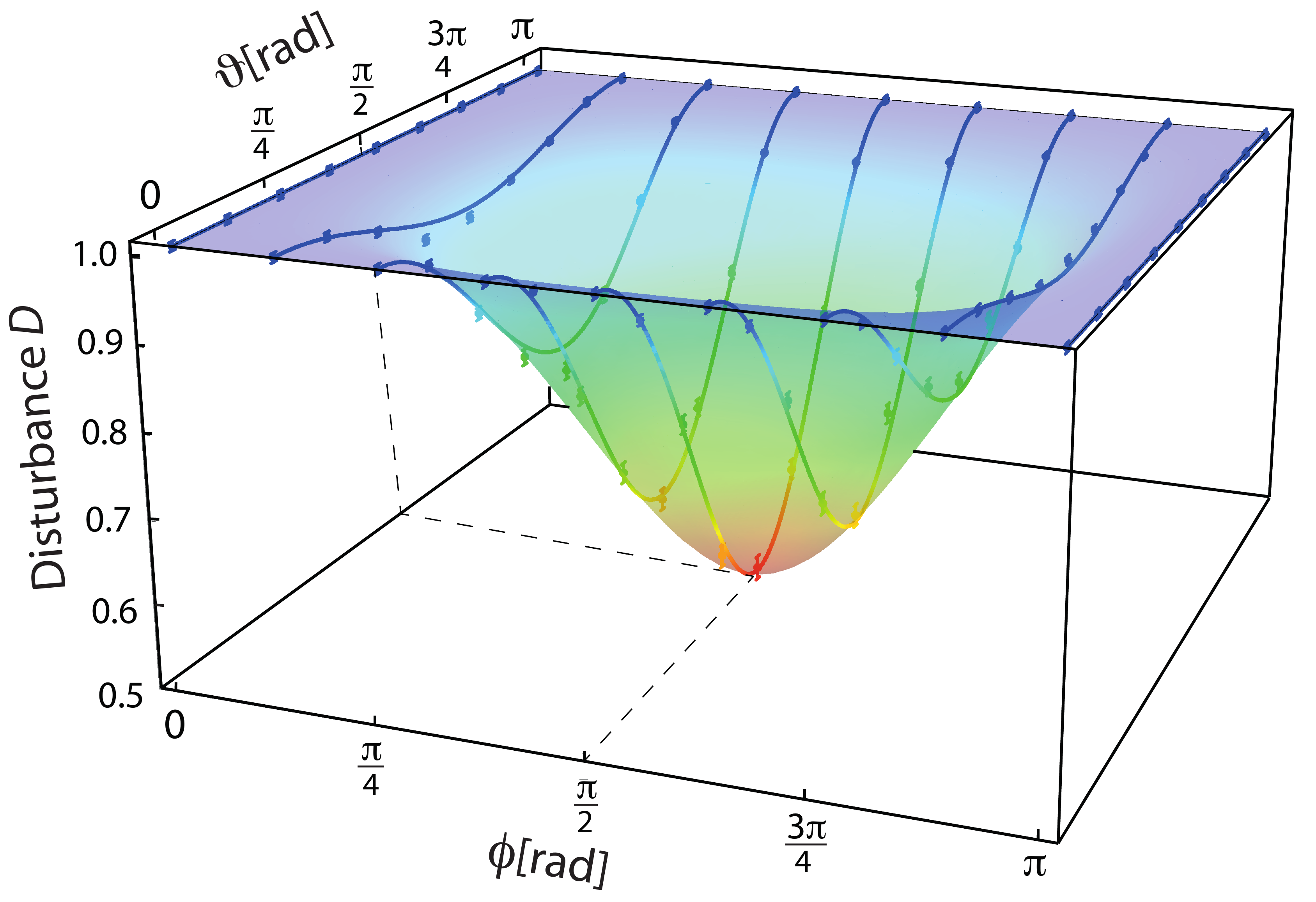}
	\caption{(color online) Disturbance data points and their respective theory curves if, after the projective measurement of $M$ (with $\theta$=$50\degree$) the eigenstate $\ket{+ m}$ is rotated onto states $\ket{\psi(\vartheta,\phi)}=\cos\frac{\vartheta}{2}\ket{+z}+e^{i\phi}\sin\frac{\vartheta}{2}\ket{-z}$ and $\ket{- m}$ onto $\ket{-\psi}=\ket{\psi(\pi-\vartheta,\phi+\pi)}$.}
	\label{fig:exp_correction_search}
\end{figure}

The disturbance induced on $B=\sigma_y$ by the measurement of $M$ behaves reciprocally to the noise in Fig.\,\ref{fig:NoiseDistUnCorr}, exemplifying the trade-off between noise and disturbance in Eq.\,(\ref{eq:ndur}). If $M=\sigma_z$ the result of a subsequent $B$-measurement allows no inference of the initially sent eigenstate $\ket{\beta y}$. The outcomes $\beta'=\pm 1$ are equally probable for both eigenstates and the information-theoretic disturbance becomes maximal. If $M=\sigma_y$ (for $\theta=90\degree$) the outcomes of the $B$-measurement are perfectly correlated with the input eigenstates. The prior measurement of $M$ leads to no loss of correlation between the $\beta$ and the measurement outcome $\beta'$ and the information theoretic disturbance vanishes. For $\theta=180\degree$ ($M=-\sigma_z$), the correlation is lost entirely, we again have $p(\beta|\beta')= \frac{1}{2}, \, \forall \beta,\beta'$ as for $\theta=0\degree$ and therefore maximal disturbance. 

\emph{Optimal correction procedure-} For the special cases $M=\pm A$ and $M =\pm B$ the disturbance is fixed to be either 1 or 0, but for the intermediate values, it can be reduced by performing a correction operation $\mathcal{C}$ after the $M$-measurement. An important class of correction operations are unitary transformations which could be experimentally realized by an additional spin turner device. However, we can concatenate state preparation after the $M$-measurement and correction $\mathcal{C}$ and immediately prepare the rotated state with DC-3 alone. In Fig\,.\ref{fig:exp_correction_search}, we show experimental results if, after the projective measurement of $M$ with $\theta=50\degree$, the eigenstates $\ket{\mu m}$ are rotated along arbitrary directions, that is, onto states $\ket{\psi(\vartheta,\phi)}=\cos\frac{\vartheta}{2}\ket{+z}+e^{i\phi}\sin\frac{\vartheta}{2}\ket{-z}$ for $\ket{+m}$ and $\ket{-\psi}=\ket{\psi(\pi-\vartheta,\phi+\pi)}$ for $\ket{-m}$. The directions of the output states are varied over the region $[\vartheta \times \phi] = [[0\degree,180\degree]\times [0\degree,180\degree]]$ with step widths $\Delta \vartheta=\Delta\phi=22.5\degree$. The minimal disturbance is obtained if the eigenstates $\ket{\pm m}$ are rotated exactly onto the eigenstates $\ket{\pm y}$ of $B$, that is for $\vartheta=\phi=90\degree$.

\begin{figure}[!b]
	\includegraphics[width=67mm]{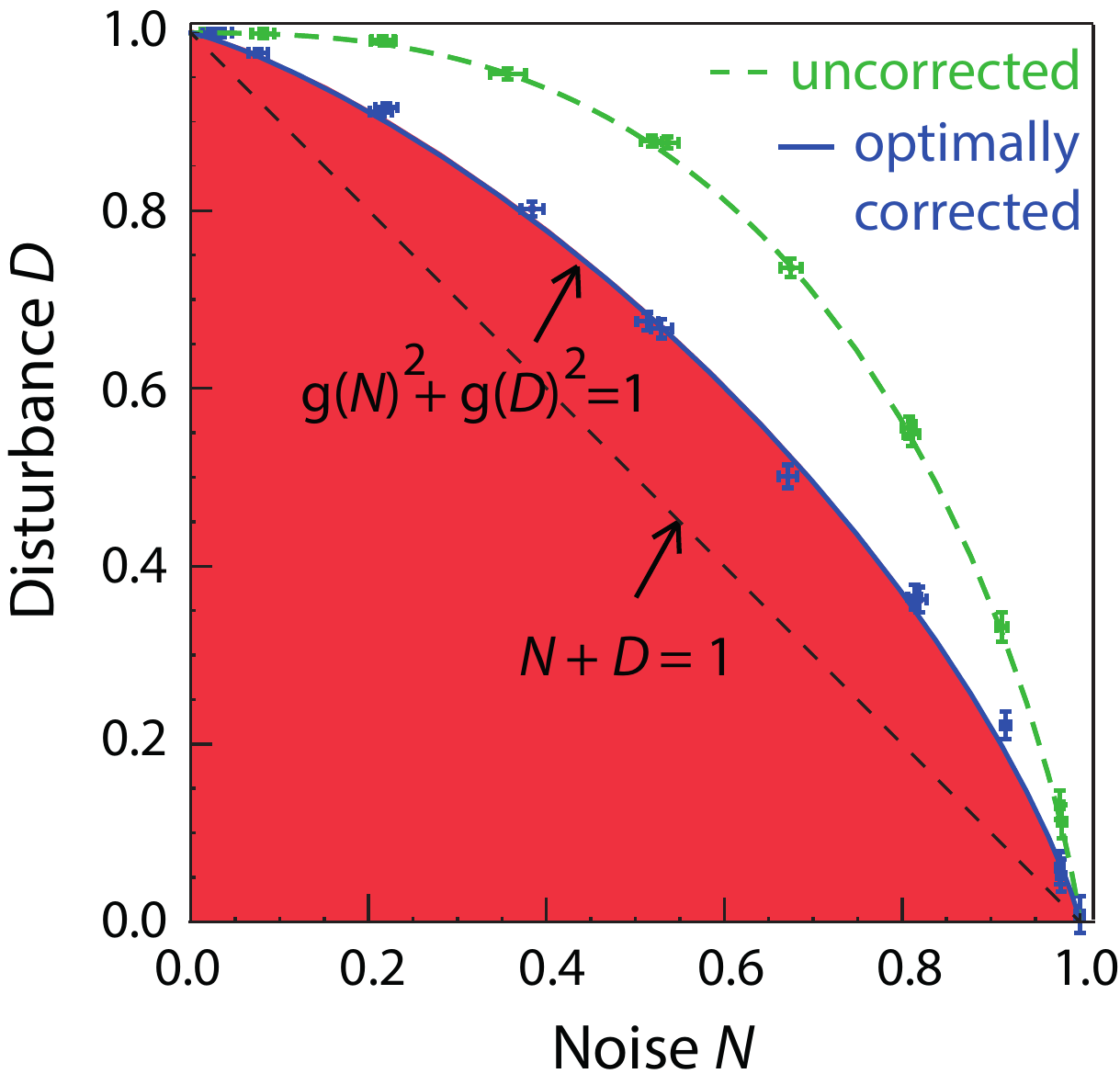}
	\caption{(color online) Disturbance vs. Noise with and without optimal correction procedure. The red shaded area marks the region which are prohibited according to Eq.\,(\ref{eq:qubits_ndur}).}
	\label{fig:NoiseDisturbancePlane}
\end{figure}

This experimental result can be generalized for the projective measurement of $M=\vec m \cdot \vec{\sigma}$ on observable $B =\vec{b}\cdot\vec{\sigma}$ yielding the optimal error correction ${\cal C}_{\rm opt}$
	\begin{equation}
	{\cal C}_{\rm opt}(|\mu {m}\rangle) := 
	\begin{cases}
	|\mu {b}\rangle,&\vec{b}\cdot \vec{m}\geq0\\
	|-\mu {b}\rangle,&\vec{b}\cdot \vec{m}<0
	\end{cases}
	\end{equation}
with $\mu =\pm 1$ and $|\pm {m}\rangle$,$|\pm {b}\rangle$ being the respective eigenstates of $M$ and $B$ (see Sec.\,IV of \cite{supp_entropic_PRL} for a detailed explanation and proof).
The results for the optimally corrected disturbance are depicted in Fig.\,\ref{fig:NoiseDistUnCorr}, together with its theoretically expected curve given, using $h(x)$ from Eq.\,(\ref{eq:h_def}), by
\begin{equation}
\label{eq:dist_minimized}
D_{\rm opt}(\mathcal{M}, B)= h\left(\sin\theta\right).
\end{equation}

In order to investigate the uncertainty relations Eqs. (\ref{eq:ndur}) and (\ref{eq:qubits_ndur}) we plot disturbance and noise data pairs from Fig.\,{\ref{fig:NoiseDistUnCorr}} against each other in Fig.\,\ref{fig:NoiseDisturbancePlane}. We immediately see that the noise disturbance uncertainty relation Eq.\,(\ref{eq:ndur}) is always fulfilled, but not saturable apart from extremal values, that is when either $N$ or $D$ vanishes. In contrast, the improved qubit relation Eq.\,(\ref{eq:qubits_ndur}) provides a tight bound and can be saturated if the optimal correction procedure is applied, as in our experiment.

\emph{Conclusions -} 
We have shown the experimental validity of the information-theoretic formulation of Heisenberg's noise-disturbance uncertainty principle in qubit measurements. As soon as we obtain knowledge about the value of a certain spin observable by applying a suitable measurement the information which can be extracted about another, incompatible observable is reduced. For maximally incompatible spin observables we observe a completely reciprocal trade-off. In order to characterize the irreversible loss of correlations, correction operations are performed which minimize the disturbance, but the sum of noise and disturbance is always bounded from below. We mathematically characterized and experimentally confirmed the optimal correction procedure for qubits, leading to a tight noise-disturbance uncertainty relation. This result should stimulate the search for improved entropic uncertainty relations for observables of higher dimensional Hilbert spaces as well.

\begin{acknowledgments}
\emph{Acknowledgements -} This work was supported by the Austrian science fund (FWF) projects P24973-N20 and P25795-N20. F.B announces support from the JSPS KAKENHI, No.~26247016. M. J.W. H. is supported by the ARC Centre of Excellence CE110001027. M. O. acknowledges support from the John Templeton Foundations, ID 35771, JSPS KAKENHI No.~26247016, and MIC SCOPE No.~121806010.
\end{acknowledgments}

\newpage
 
\onecolumngrid
 
\appendix
 
\section{Supplementary Material}

\subsection{Improved information-theoretic uncertainty relation for qubits}
\label{sec:improved_inequality}

A generally valid uncertainty relation between noise $N(\mathcal M,A)$ and disturbance $D(\mathcal M,B)$, for any measurement apparatus $M$, is given by Eq.\,(4) of the main text---which for our investigated qubit scenario reduces to 
\begin{equation} 
\label{ndur}
N(\mathcal M,\sigma_z) + D(\mathcal M,\sigma_y) \geq \log 2 = 1~{\rm bit}.
\end{equation}
For our experiment, this inequality is only saturated for extremal values, that is, if either $N$ or $D$ vanish. By applying the optimal correction operation for rank-one projective measurements, the measured values come closer to the straight line $N+D=1$ in the $N$-$D$ plane, but do not reach it (see blue data points in Fig.\,5 of the main text). Thus, the question arises as to whether the above inequality can be saturated by a different class of measurements, or, conversely whether an improved, saturable inequality exists.

Here we show that the above inequality can in fact be substantially improved, to
\begin{equation} \label{opt}
g[ N(\M,\sigma_z)]^2 + g[ D(\mathcal M,\sigma_y)]^2 \geq 1,
\end{equation}
where $g[x]$ is the inverse of the function 
$h(x) := -\frac{1+x}{2} \log_2 \frac{1+x}{2} - \frac{1-x}{2}\log_2 \frac{1-x}{2}$
on the interval $x\in [0,1]$.  This inequality is in fact {\it optimal}, i.e., it is the tightest possible inequality for the noise and disturbance of $\sigma_z$ and $\sigma_y$, for arbitrary measurement apparatuses.  Moreover, this optimal inequality is saturated in our experimental scenario, as depicted in Fig.\,5 of the main text.

To prove Eq.\,(\ref{opt}), let $R:=\{(N,D)\}$ denote the region of possible values of $N=N(\M,\sigma_z)$ and $D=D(\mathcal M,\sigma_y)$, over all possible measurements $\M$. Hence, $R$ has some lower boundary, $C$, that in general prevents the noise and disturbance from both being arbitrarily small.  Indeed, from Eq.\,(\ref{ndur}) above, all points in $C$ must lie on or above the line $N+D=1$ bit.  Our aim is to show that $C$ is given by Eq.\,(\ref{opt}).

Now, as shown in the Supplemental Material of \cite{Buscemi14}, the  noise and disturbance of two system observables $A$ and $B$, for any measurement $M$, satisfy
\[
N(\M,A) \geq \sum_u p_u H(A^T|\rho_u)=\sum_u p_u H(A|\rho_u^T),~~~~~
D(\M,B) \geq \sum_u p_u H(B^T|\rho_u)=\sum_u p_u H(B|\rho_u^T), 
\]
where $\E=\{ \rho_u;p_u\}$ is an ensemble of reduced states describing the system $S$, following measurement of some POVM $\{\Pi_u\}$ on a reference copy $R$ of the system that is maximally entangled with $S$.  Here $H(C|\rho)$ denotes the entropy of $C$ for state $\rho$, and the transpose $C^T$ of operator $C$ is defined with respect to the Schmidt basis of the maximally entangled state of $R$ and $S$ \cite{Buscemi14}. 

It immediately follows that $C$, the lower boundary of $R$,  lies on or above the lower boundary $C^*$ of the region
\begin{equation}
R^*:= \left\{ \left( \sum_m p_m H(\sigma_z|\rho_m),\sum_m p_m H(\sigma_y|\rho_m)\right) \right\}, 
\end{equation}
where $\E=\{\rho_m;p_m\}$ ranges over all possible ensembles of qubit states (thus including the ensembles $\{\rho^T_u;p_u\}$ in particular). Remarkably, it turns out that $C\equiv C^*$, i.e., $C^*$ specifies the optimal uncertainty relation for  $N(\M,\sigma_z)$ and $ D(\M,\sigma_y)$ .

The explicit form of the curve $C^*$ may be determined by showing that attention can be restricted to the subset of pure-state ensembles, and performing a suitable variational calculation. First, for a given ensemble $\E=\{\rho_m;p_m\}$, let $r^{(m)}$ denote the Bloch vector corresponding to the qubit state $\rho_m$, i.e., $\rho_m=\frac{1}{2}(1+\sigma\cdot r^{(m)})$.  Now define a corresponding {\it pure}-state ensemble, $\E':=\{\rho'_m;p_m\}$, by taking $r^{(m)'}$ to be the unit vector in the $y$-$z$~plane which has (i) the same component as $r^{(m)}$ in the $y$-direction; (ii) no component in the $x$-direction;  and (iii) a remaining component in the $\pm z$-direction, with the $+$ sign ($-$ sign) chosen if  $r^{(m)}\cdot z\geq 0$ ($<0$). Thus, $|r^{(m)'}|=1$, $r^{(m)'}\cdot y=r^{(m)}\cdot y$, and $|r^{(m)'}\cdot z|\geq |r^{(m)}\cdot z|$ (since, by construction, $r^{(m)'}$ has a longer component than $r^{(m)}$ in the $z$-direction).  It follows immediately that $H(\sigma_z|\rho'_m)\leq H(\sigma_z|\rho_m)$ and $H(\sigma_y|\rho'_m)=H(\sigma_y|\rho_m)$.  Hence, 
\[	\sum_m p_m H(\sigma_z|\rho'_m) \leq \sum_m p_m H(\sigma_z|\rho_m) ,~~~~~~~~\sum_m p_m H(\sigma_y|\rho'_m) = \sum_m p_m H(\sigma_y|\rho_m) . \]
Thus, for any point $(N,D)\in R^*$ generated by some ensemble $\E$, there is a point $(N',D)\in R^*$ generated by a corresponding pure-state ensemble $\E'$, with $N'\leq N$.

Hence, to find the point $(N,D)$ on the lower boundary $C^*$ of $R^*$, for any given value of $D$, one only has to minimise the variational quantity
\begin{eqnarray} \nonumber
J &=& \sum_m p_m H(\sigma_z|\rho'_m) + \kappa \left[ \sum_m p_m H(\sigma_y|\rho'_m) - D\right] +\lambda \left[\sum_m p_m - 1\right] \\
&=& \sum_m p_m \,h(\sin\theta_m) + \kappa \left[ \sum_m p_m \,h(\cos\theta_m) - D\right] +\lambda \left[\sum_m p_m - 1\right] \label{var}
\end{eqnarray}
over all pure-state ensembles $\E'=\{\rho'_m;p_m\}$ in the $yz$-plane.  Here $\kappa$ and $\lambda$ are Lagrange multipliers, the Bloch vector is parameterised as $r^{(m)'}=(0, \cos\theta_m,\sin\theta_m)$, and $h(x)$ is defined as above.

\begin{figure}[!t]
	\centering
	\includegraphics[width=0.75\textwidth]{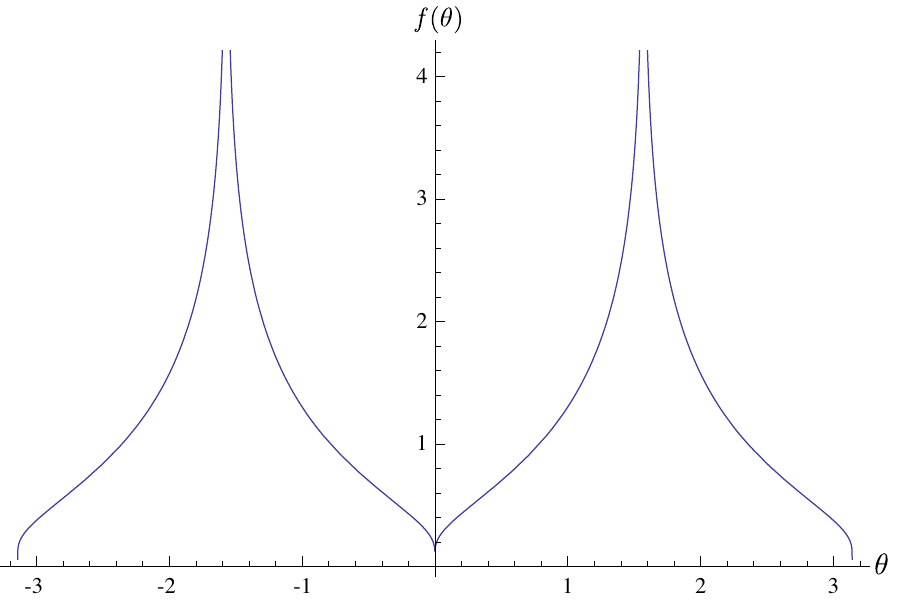}
	\caption{ The function $f(\theta)$ in Eq.\,(\ref{kapconqd}).}
	\label{fig:fqd}
\end{figure}

There are two sets of variational equations, for $\{p_m\}$ and $\{\theta_m\}$ respectively, which fix the Lagrange multipliers $\lambda$ and $\kappa$.  In particular,  $\partial J/p_m=0$ yields
\begin{equation} \label{lamconqd}
h(\sin\theta_m) + \kappa \,h(\cos\theta_m) = - \lambda ,
\end{equation}
while  $\partial J/\theta_m=0$ yields
\[ p_m \big[ \cos\theta_m h'(\sin\theta_m) -\kappa \sin\theta_m h'(\cos\theta_m) \big] =0. \]
The latter reduces to, for all $p_m\neq0$ (i.e, for those $p_m$ which actually contribute to $J$) 
\begin{equation} \label{kapconqd}
f(\theta_m):=\frac{ h'(\sin\theta_m)/\sin\theta_m}{h'(\cos\theta_m)/\cos\theta_m} =  \kappa .
\end{equation}
It may be checked (see Fig.\,\ref{fig:fqd}) that the function $f(\theta)$ in Eq.\,(\ref{kapconqd}) is symmetric about $\theta=0$ and $\theta=\pi/2$, and monotonic on $[0,\pi/2]$.   Hence, one has up to four possible solutions of $f(\theta_m)=\kappa$, for a given value of $\kappa$, of the form $\theta_m=\pm\pi/2\pm\theta$ for some $\theta\in[0,\pi/2]$.  Thus,
\begin{equation} \nonumber
\sin\theta_m = \pm \cos \theta,~~~~~~\cos\theta_m=\mp \sin\theta,~~~~~\forall m: p_m\neq 0,
\end{equation}
for some fixed $\theta\in[0,\pi/2]$, and $\kappa=f(\pi/2-\theta)$. Moreover, noting $h(x)=h(-x)$, Eq.\,(\ref{lamconqd}) is automatically satisfied, with $\lambda=-h(\cos\theta)-\kappa \,h(\sin(\theta)$.

The corresponding extremal value of $J$ follows from Eq.\,(\ref{var}) as
$J_{\rm min} = h(\cos\theta)$, independently of $p_m$, yielding the lower boundary of the region $R^*$ to be
\begin{equation} \label{cabsolqd}
C^* ={\large\{ }\left(h(\cos\theta), h(\sin\theta)\right): \theta\in [0,\pi/2] \,{\large\}} .
\end{equation}
Noting that $\cos^2\theta+\sin^2\theta=1$, the equation of this curve corresponds to equality in Eq.\,(\ref{opt}).

Finally, to show that $C^*\equiv C$, note that all the points $(N,D)\in C^*$ correspond to the values of noise and disturbance for the measurements in our experimental scenario (with optimal correction), as per Eqs.\,(10) and~(11) of the main text.  Hence, $C^*\subset R$, implying $C$ lies on or below $C^*$.  But,  by construction, $C^*$ lies on or below $C$, and it immediately follows that $C=C^*$ as desired.

Similar techniques to the above may be used to derived improved noise-disturbance relations for other observables, as will be explored elsewhere.  We note here that, by restricting attention to single-member ensembles in Eq.\,(\ref{var}), it further follows that $C^*$  is the lower boundary of the region $\{ H(\sigma_z|\rho), H(\sigma_y|\rho)\}$, where $\rho$ ranges over all possible qubit states $\rho$.  Hence, $C^*$ also gives the optimal trade-off between the entropies of the mutually complementary qubit observables $\sigma_y$ and $\sigma_z$, improving on the standard Maassen-Uffink relation $H(\sigma_y)+H(\sigma_z)\geq \log 2$.

\subsection{State preparation and measurement in the neutron polarimetric setup}
\label{sec:state}

Here, the experimentally interested reader can find more detailed informations on how the neutron's spin state is manipulated along our polarimeter beam line in order to prepare the desired initial states ($\ket{\pm z}, \ket{\pm y}$) and successively measure the observables of interest $M= \sigma_y \sin\theta  + \sigma_z \cos\theta$ and $B=\sigma_y$.

\subsubsection{State preparation}
In order to prepare the eigenstates $\ket{\alpha z}$ of $A$, that is $\ket{+ z}$ and $\ket{-z}$ since $\alpha=\pm 1$, the current through the spin turner coil DC-1 is simply turned off for the former leaving the the spin in the state it possesses after the first supermirror, i.e. $\ket{+ z}$, and set to the predetermined  flip current generating $B_x^{\pi}$ for the latter.  $B_x^{\pi}$ is just the field strength that causes a rotation of $\pi$ of the Bloch vector and thus converts $\ket{+ z}$ to $\ket{-z}$. To prepare the eigenstates $\ket{\beta y}$ of B, the respective currents generating the fields $B_x^{\pm\pi/2}$, i.e., that cause $\pm \pi/2$-roations of the Bloch vector, are applied in DC-1. Each of the two eigenstates of $\sigma_z$ and $\sigma_y$ is sent with equal probability, experimentally realized by sending each input state for the same, sufficiently long time period.

\subsubsection{Measurement of $M$}
The projective measurement of $M$ consists of two steps. At first, we have to project the initially prepared state onto the eigenstates of $M$, then, in order to complete the measurement, we have to prepare the neutron spin in the eigenstates of $M$. Since the input eigenstates $\ket{\pm z}$ and $\ket{\pm y}$ and the $\ket{\pm{ m}}$ all lie in the $zy$-plane of the Bloch sphere, the distance between DC-1 and DC-2 has to be chosen such that the Bloch vector undergoes integer multiples of the full rotation period in their intermediate guide field. Then, spin turner DC-2 rotates the spin component to be measured, which depends on the polar angle $\theta$ ($M= \sigma_y \sin\theta  + \sigma_z \cos\theta)$, towards the z-direction. For the eigenstate belonging to eigenvalue $\mu=+1$ the component along $+\vec m$ and for eigenvalue $\mu=-1$ the spin component along $-\vec m\ $ is rotated in the $+z$ direction.
The second supermirror (first analyzer) then selects only the $\ket{+z}$ part of the spinor wave function. The projective measurement is completed by the preparation of the measured spin component with spin turner DC-3. In analogous manner to the preparation of the initial state, this is accomplished by properly setting the respective currents in DC-3 required for the fields $B_x^{\theta}$ and $B_x^{\pi+\theta}$. Thus when leaving DC-3 the system is in the appropriate eigenstate of $M$. 

\subsubsection{Measurement of $B$}
Here we apply the same procedure as for the $M$ measurement, that is to rotate the $\pm y$ component towards the $+z$-direction with DC-4, followed by another supermirror (second analyzer). A further DC coil for preparing the measured spin state can be omitted, since the neutron detection is insensitive to the spin.

\subsection{From intensities to noise and disturbance}
\label{sec:from}

In this section, we want to explain in detail how the probabilities needed for the calculation of noise and disturbance are obtained from the intensities measured in the experiment.

\subsubsection{Noise}

In order to determine the information-theoretic noise, the eigenstates $\ket{\pm z}$ of $A$ are sent onto the measurement apparatus which then projectively measures $M$ and $B$ resulting in four different output intensities for each input eigenstate. We have schematically depicted the measurement process in Fig.\,\ref{fig:joint_apparatus_intensities_noise}. The polarimeter setup is adjusted such that it realizes one of the eight possible "arms" of Fig.\,\ref{fig:joint_apparatus_intensities_noise} after the other. 
\begin{figure*}[!htb]
  \includegraphics[width=125mm]{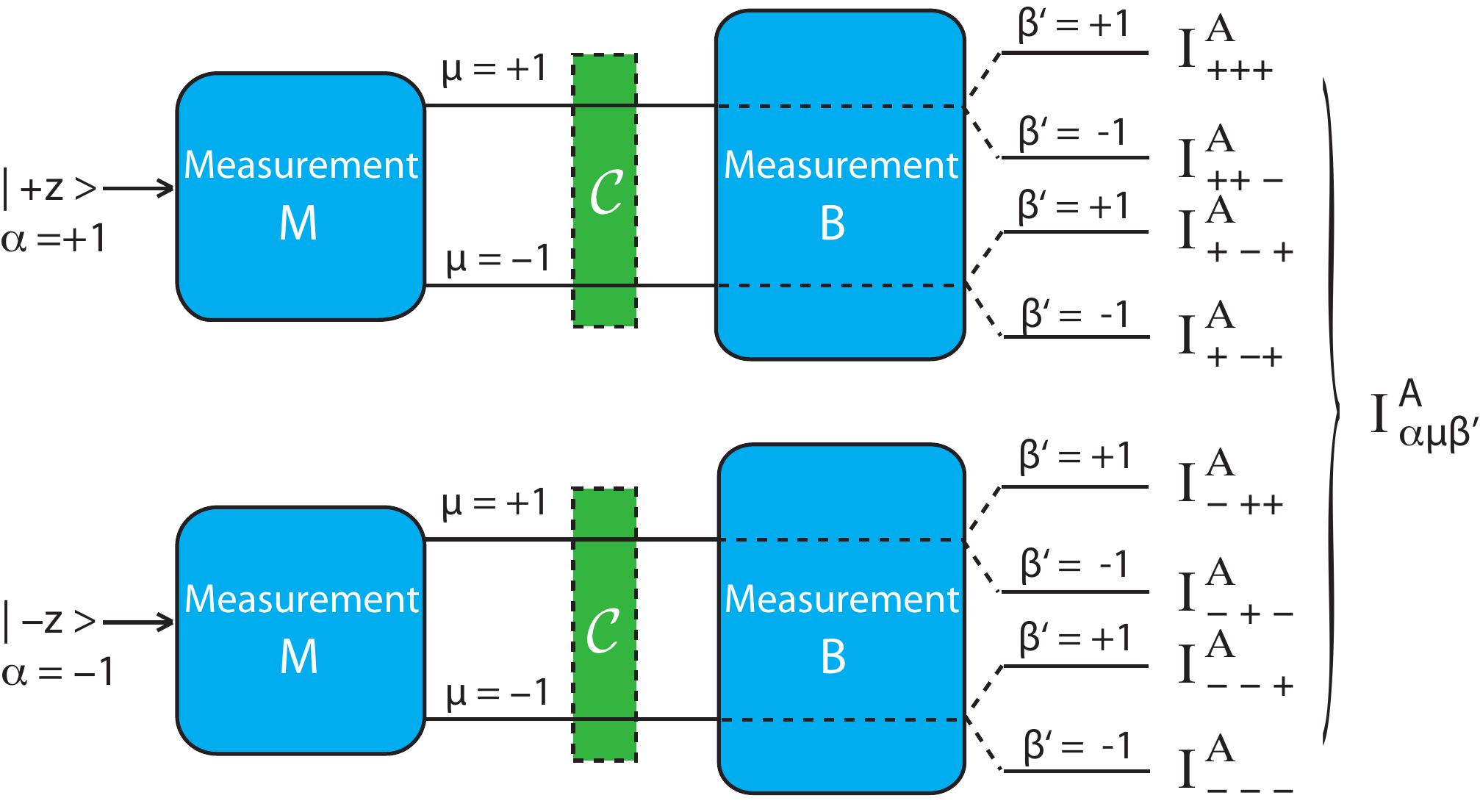}
  \caption{Schematic illustration of the complete noise measurement procedure. All probabilities required for the determination of the information-theoretic noise can be obtained from the output intensities $I_{\alpha \mu\beta'}^{A}$.  
}
\label{fig:joint_apparatus_intensities_noise}
 \end{figure*}
The output intensities get labeled with three lower indices $I^{A}_{\alpha \mu \beta'}$ having the values $\pm 1$ where $\alpha$ gives the sign of $\ket{\pm z}$, $\mu$ indicates which projection operator of $M$ has been realized, and $\beta'$ does the same for the projection operator of $B$. The probabilities are connected to the intensities via 
\begin{equation}
\label{eq:prob_noise_exp}
p(\alpha)=\frac{\sum_{\mu,\beta'} I^{A}_{\alpha \mu \beta'}}{\sum_{\alpha,\mu,\beta'} I^{A}_{\alpha \mu \beta'}}  \qquad 
p(\mu|\alpha) = \frac{\sum_{\beta'} I^{A}_{\alpha \mu \beta'}}{\sum_{\mu,\beta'} I^{A}_{\alpha \mu \beta'}}.
\end{equation}  
However, the definition of the information theoretic noise is not in terms of the conditioned probability $p(\mu|\alpha)$ but $p(\alpha|\mu)$, since it quantifies how well the observable's value can be guessed from the outcome and not contrariwise:
\begin{equation}
\label{eq:noise_definition}
N(\mathcal M, A) := - \sum_{\alpha,\mu} p(\mu) p(\alpha|\mu) \log p(\alpha|\mu)= H(\mathbb{A}|\mathbb{M})
\end{equation}
Here $H(\mathbb{A}|\mathbb{M})$ is the conditional entropy and $\mathbb{A}$ and $\mathbb{M}$ denote the classical random variables associated with input $\alpha$ and output $\mu$. The information-theoretic noise thus quantifies how well the value of $A$ can be inferred from the measurement outcome and only vanishes if an absolutely correct guess is possible.
Thus, we have to use Bayes' theorem to connect these different conditional probabilities
\begin{equation}
\label{eq:bayes_law}
p(\alpha|\mu)=\frac{p(\alpha) p(\mu|\alpha)}{p(\mu)}.
\end{equation}
The marginal probability $p(\mu)$ is given by summation over $\alpha$ of the joint probability distribution $p(\alpha,\mu)$
\begin{equation}
\label{eq:pmu}
p(\mu) = \sum_{\alpha} p(\alpha,\mu) = \sum_{\alpha} p(\alpha) p(\mu|\alpha).
\end{equation}
Now, the information theoretic noise as defined in Eq.\,(\ref{eq:noise_definition}) can be calculated. 

\subsubsection{Disturbance}
For the determination of the information-theoretic disturbance, the eigenstates $\ket{\beta y}$, that is $\ket{\pm y}$, of the disturbed observable $B=\sigma_y$ are sent onto the apparatus. By labeling the output intensities with $I^{B}_{\beta\mu\beta'}$ (see Fig.\,\ref{fig:joint_apparatus_intensities_dist}) we get the required probabilities from
\begin{equation}
\label{eq:prob_dist_exp}
p(\beta)=\frac{\sum_{\mu,\beta'} I^{B}_{\beta \mu \beta'}}{\sum_{\beta,\mu,\beta'} I^{B}_{\beta \mu \beta'}} \qquad 
p(\beta'|\beta) = \frac{\sum_{\mu} I^{B}_{\beta \mu \beta'}}{\sum_{\mu,\beta'} I^{B}_{\beta \mu \beta'}}. 
\end{equation}
By again using Bayes theorem we obtain the probabilities as they occur in the definition of the information-theoretic disturbance.
\begin{equation}
\label{eq:disturbance_definition}
D(\mathcal M, B) := - \sum_{\beta,\beta'} p(\beta') p(\beta|\beta') \log p(\beta|\beta')= H(\mathbb{B}|\mathbb{B}').
\end{equation}
as given in the main text, with 
\begin{equation}
\label{eq:h_def}
h(x):=-\frac{1+x}{2}\log \frac{1+x}{2} - \frac{1-x}{2}\log \frac{1-x}{2}.
\end{equation}
\begin{figure*}[!b]
	\includegraphics[width=125mm]{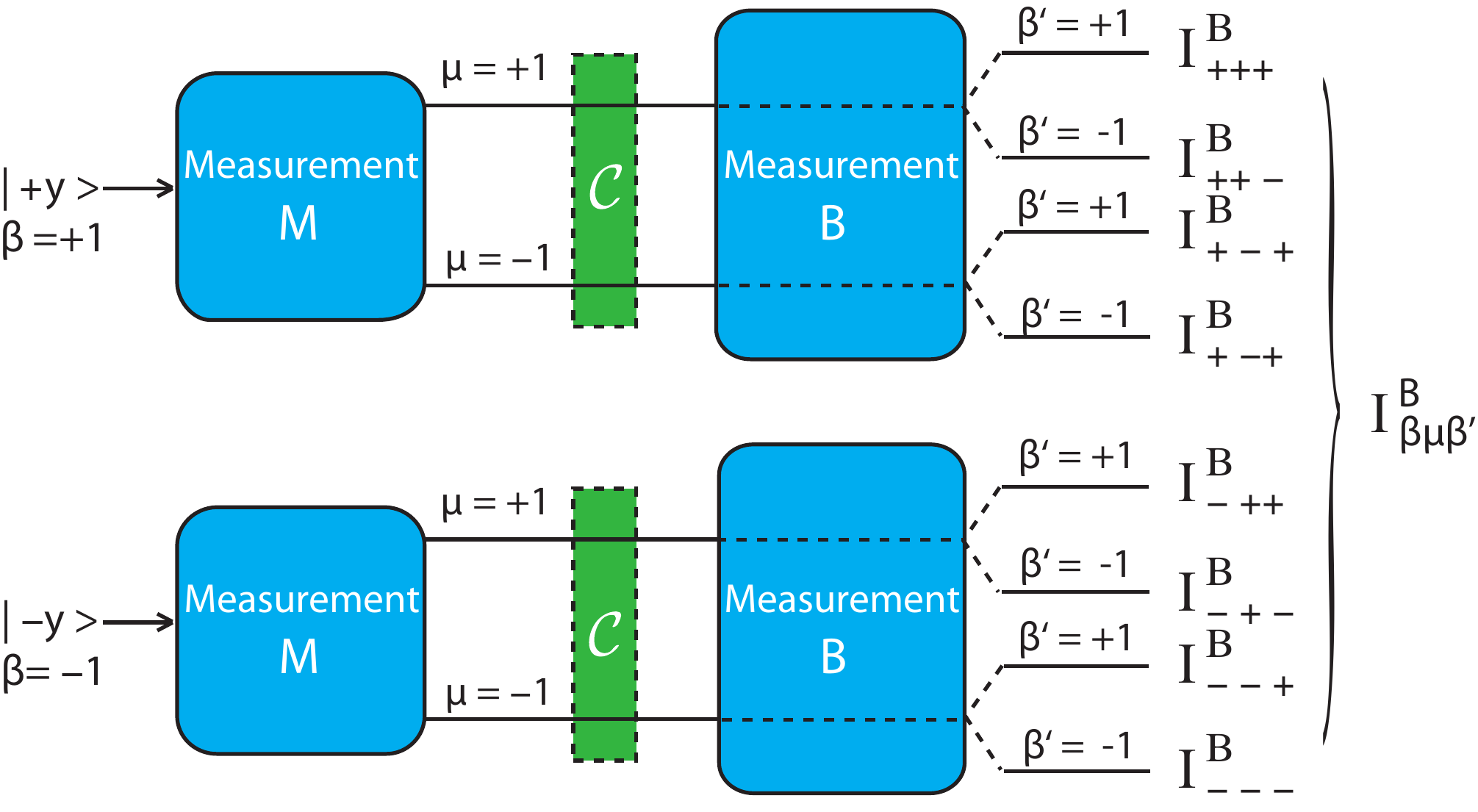}
	\caption{Schematic illustration of the complete disturbance measurement procedure. All probabilities required for the determination of the information-theoretic disturbance can be obtained from the output intensities $I_{\beta \mu\beta'}^{B}$.  
	}
	\label{fig:joint_apparatus_intensities_dist}
\end{figure*}
In our scenario, the measurement operator $M$ is varied over the zy-plane spanned by $A$ and $B$
\begin{equation}
	\label{eq:ABM}
A=\sigma_z, \quad B=\sigma_y,\quad	M=\vec m\cdot\vec{\sigma} = \sigma_y \sin\theta + \sigma_z\cos\theta,
\end{equation}
and the theoretically expected expressions for the probabilities are 
\begin{eqnarray}
\label{eq:prob_noise_theory}
p(\alpha)= p(\beta)= p_{\rm opt}(\beta)=\frac{1}{2},  \qquad 
p(\mu|\alpha)& =& \frac{1+ \mu\alpha\cos\theta}{2},   \nonumber\\ 
p(\beta'|\beta) =p(\beta|\beta') = \frac{1+ \beta'\beta\sin^2\theta}{2}, \quad 
p_{\rm opt}(\beta'|\beta) &=&p_{\rm opt}(\beta|\beta') = \frac{1+ \beta'\beta |\sin\theta|}{2}.
\end{eqnarray}
Here $p_{\rm opt}$ is used to denote the probabilities in the case that  an optimal correction (see section\,\ref{sec:optimal}) is applied after the $M$-measurement. These expressions yield 
\begin{equation}
\label{eq:noise_dist_qbit}
N(\mathcal M,A) = h\left(\cos\theta\right), \quad 
D_0(\mathcal M,B) =  h\left(\sin^2\theta\right), \quad 
D_{{\rm opt}}(\mathcal M,B) =  h\left(\sin\theta\right),
\end{equation}

These theoretically expected value of noise and disturbance are depicted here in Fig.\,\ref{fig:NoiseDistUnCorr}, along with the corresponding experimentally measured values.
\begin{figure}[!htb]
	\includegraphics[width=170mm]{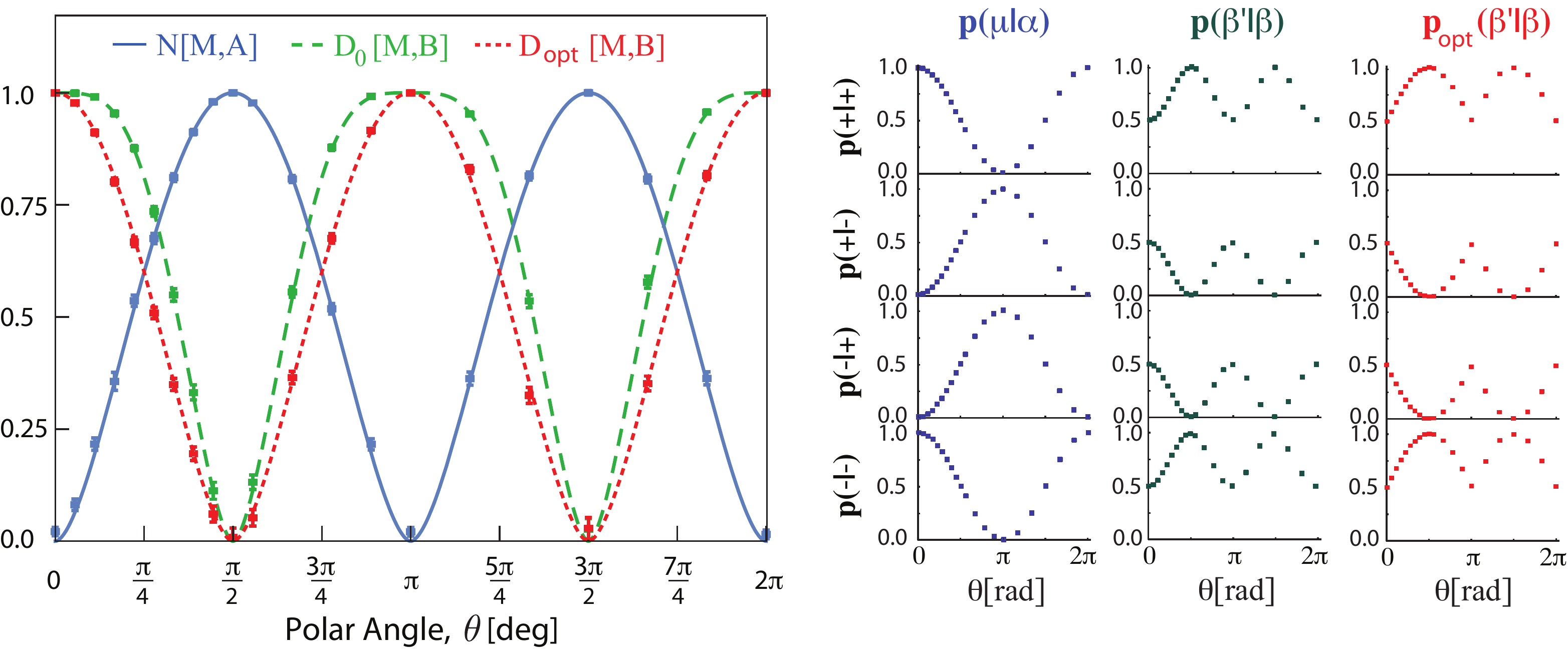}
	\caption{Noise $N(M,A)$ (straight blue line), uncorrected disturbance $D_0(M,B)$ (dashed green line), and  optimally corrected disturbance $D_{{\rm opt}}(M,B)$ (dotted red line) normalized with the apparatus' efficiency vs. polar angle $\theta$ of $M$. The respective conditional probabilities are plotted on the right hand side. }
	\label{fig:NoiseDistUnCorr}
\end{figure}

\subsection{Optimal correction procedure for projective qubit measurements}
\label{sec:optimal}
For a spin measurement operator $M=\vec m \cdot \vec{\sigma}$ and an observable $B =\vec{b}\cdot\vec{\sigma}$ the optimal correction minimizing the disturbance after the projective measurement of $M$ on $B$ is given by
\begin{equation}
{\cal C}_{\rm opt}(|\mu {m}\rangle) := 
\begin{cases}
|\mu {b}\rangle,&\vec{b}\cdot \vec{m}\geq0\\
|-\mu {b}\rangle,&\vec{b}\cdot \vec{m}<0
\end{cases}
\end{equation}
with $\mu=\pm 1$ and $|\pm {m}\rangle$ being the respective eigenstates of $M$. The above formula can be intuitively understood as a sort of \textit{maximum likelihood} correction procedure: the output $|\mu{m}\rangle$ of the apparatus is rotated onto the closest eigenvector of $B$, which is $|\mu {b}\rangle$, if $M$ and $B$ are more correlated than anti-correlated (i.e., $\vec{b}\cdot\vec{m}\ge 0$), or $|-\mu{b}\rangle$ otherwise (i.e., $\vec{b}\cdot\vec{m}< 0$). This guarantees that the outcome obtained from the final measurement of $B$ is perfectly correlated with the outcome of $M$, if $\vec{b}\cdot\vec{m}\ge 0$, or perfectly anti-correlated otherwise---in either cases, correlations are kept maximal, by avoiding the occurrence of extra random noise.

The fact that such a simple correction procedure is indeed the optimal one, is a consequence of having an apparatus performing a sharp measurement (i.e., measurement operators are all rank-one). In such a case, in fact, the apparatus is necessarily of the form `measure-and-prepare,' in the sense that the apparatus has to completely absorb the input quantum system in order to produce the outcome. Therefore, the state of the output quantum system emerging from the apparatus only depends on the outcome, and all the information about the state of the input quantum system is encoded on the apparatus outcome only.

This in turns implies that, from an information-theoretic viewpoint, in the case of an apparatus performing a sharp measurement (as we have here), the best thing to do, in order to minimize the disturbance in Eq.\,(3) of  the main text, is to have the random output variable $\mathbb{B}'$ perfectly correlated (or perfectly anti-correlated) with the apparatus'{\tiny } random variable $\mathbb{M}$, so that all available information about the input system is encoded in $\mathbb{B}'$. Thus, the minimum possible disturbance is
\begin{equation}
D_{\rm opt}(\mathcal{M},B)=H(\mathbb{B}|\mathbb{B}')=H(\mathbb{B}|\mathbb{M}),
\end{equation}
for such measurements, corresponding to  the optimal correction operation in the main text.

\end{document}